# Detection of HIV-1 antigen based on magnetic tunnel junction sensor and magnetic nanoparticles

L. Li[1], K. Y. Mak[1], Y. Zhou[2], W.W. Wang[3], P. W. T. Pong[1]

[1]Department of Electrical and Electronic Engineering, The University of Hong Kong,
[2]Department of Physics, The University of Hong Kong, Hong Kong
[3]Department of Physics, Ningbo University, Ningbo 315211, China

**Abstract**

In recent years, it is evidenced that the individuals newly infected HIV are transmitting the virus prior to knowing their HIV status. Identifying individuals that are early in infection with HIV antibody negative (window period) remains problematic. In the newly infected individuals, HIV antigen p24 is usually present in their serum or plasma 7-10 days before the HIV antibody. After antibody production initiates, the p24 antigen is bound into immune complexes. That means the detectable p24 antigens in serum/plasma are short-lived, and their amount is in the pg/ml range. Thus, a rapid quantitative bio-detection system with high-sensitivity is required to achieve early disease diagnosis. Magnetoresistive (MR) biosensor with ultra-high sensitivity possesses great potential in this area. In this study, a p24 detection assay using MgO-based magnetic tunnel junction (MTJ) sensor and 20-nm magnetic nanoparticles is reported.

**Introduction**

The early disease diagnosis requires the detection and quantification of disease biomarkers in serum, urine, or other bloody fluids. Benefit from the recent advancement of nanotechnology and thin-film technology, the magnetic immunoassay (MIA) utilizing magnetic nanoparticles and magnetoresistive sensor have attracted increasing attention in the biomedical area, due to its high sensitivity, low power consumption, low cost, and feasibility to be miniaturized.

Magnetic tunnel junction (MTJ) sensors, one advanced kind of MR sensor, are recognized as an ideal sensor element for the biodetection due to their low cost, high sensitivity, and lab-on-chip compatibility. Compared with GMR sensors, MTJ sensors offer even higher MR ratio (such as, 604% at room temperature) and therefore potentially higher sensitivity at low magnetic field (especially for the magnetic field <1 Oe). [1, 2] Besides the magnetic field detector, the magnetic iron oxide nanoparticles (IONPs) have been a popular choice as bio-labels due to their physical and chemical stability, low toxicity, environmentally safe, and inexpensive to produce. The biodetection configuration utilizing IONPs and MTJ sensors has successfully been applied for bimolecular recognition and DNA detection in previous studies. [3-5] In this work, we demonstrate feasibility of magnetic biodetection for p24 antigens by using carboxyl-group functionalized magnetic IONPs with MTJs array sensors.

p24 antigens are used here as biotarget is due to the fact that p24 antigen is the biomarker of HIV disease. Acquired immunodeficiency syndrome (AIDS) is the end-stage disease caused by human immunodeficiency virus (HIV) infection.[6] Early diagnosis of HIV and thus early treatment of patients has been confirmed to bring dramatic survival benefits, while unfortunately the early detection of HIV remains problematic. In recent years, it is evidenced that the individuals newly infected with HIV have transmitted the virus prior to knowing their HIV status. In the newly infected individuals, HIV antigen p24 is usually present in their serum or plasma 7-10 days before the HIV antibody. The detectable amount of p24 antigens in serum/plasma is in the pg/ml range [1].Thus, the development of a rapid quantitative bio-detection system with high-sensitivity to detect the p24 antigen will greatly benefit the early diagnosis of HIV.

### 1.1. Experimental part

For the magnetic detection of p24, both of the IONPs and MTJ sensors need to be surface biologically functionalized. The magnetic iron oxide nanoparticles ($Fe_3O_4$, purchased from Ocean Nanotech) with carboxyl groups were used for the immobilization of the detecting antibodies via covalent bonding. [7] Anti-p24 antibodies were purchased from Abcam. The iron oxide nanoparticles (IONPs) were characterized under TEM observation and their magnetic property measurement was carried out at room temperature by using a vibrating sample magnetometer (Lakeshore, VSM 7400). The magnetization was measured over a range of applied field from around -10,000 to 10,000 Oe.

### 1.1.1. Preparation of biofunctionalized magnetic IONPs

The IONPs with carboxyl groups was conjugated with antibodies via 1-ethyl-3-(3-dimethylaminopropyl) carbodiimide hydrochloride (EDC)/ N-hydroxysuccinimide (NHS) coupling chemistry [8, 9], and Fig. 1 shows the chemical principle. Then the nanoparticles are incubated with detecting antibody and resuspended in storage buffer (PBS, phosphate buffered saline). The detailed procedure to carry out the biofunctionalization of IONPs with detecting antibodies is illustrated in Fig. 2. Briefly, the carboxyl groups-coated magnetic nanoparticles are activated by adding EDC and NHS. Then the nanoparticles are incubated with anti-p24 antibody and resuspended in storage buffer (PBS, phosphate buffered saline). Through this procedure, biofunctionalized IONPs were prepared with a concentration of 1.00 mg/ml, which can be stored for a few of months. All the buffers used here were purchased from Ocean Nanotech as a kit.

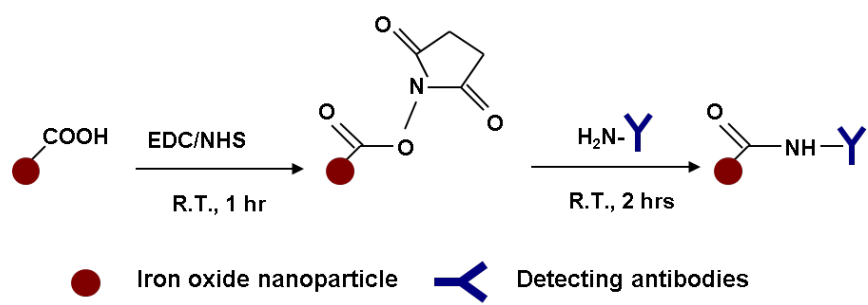

Figure 1 Schematic principle of magnetic nanoparticles functionalization with detecting antibodies: (a) introduction of carboxyl groups-coated magnetic iron oxide nanoparticles (IONPs), (b) activation of a carboxylate moiety on the surface of IONP, (c) incubation with detecting antibody

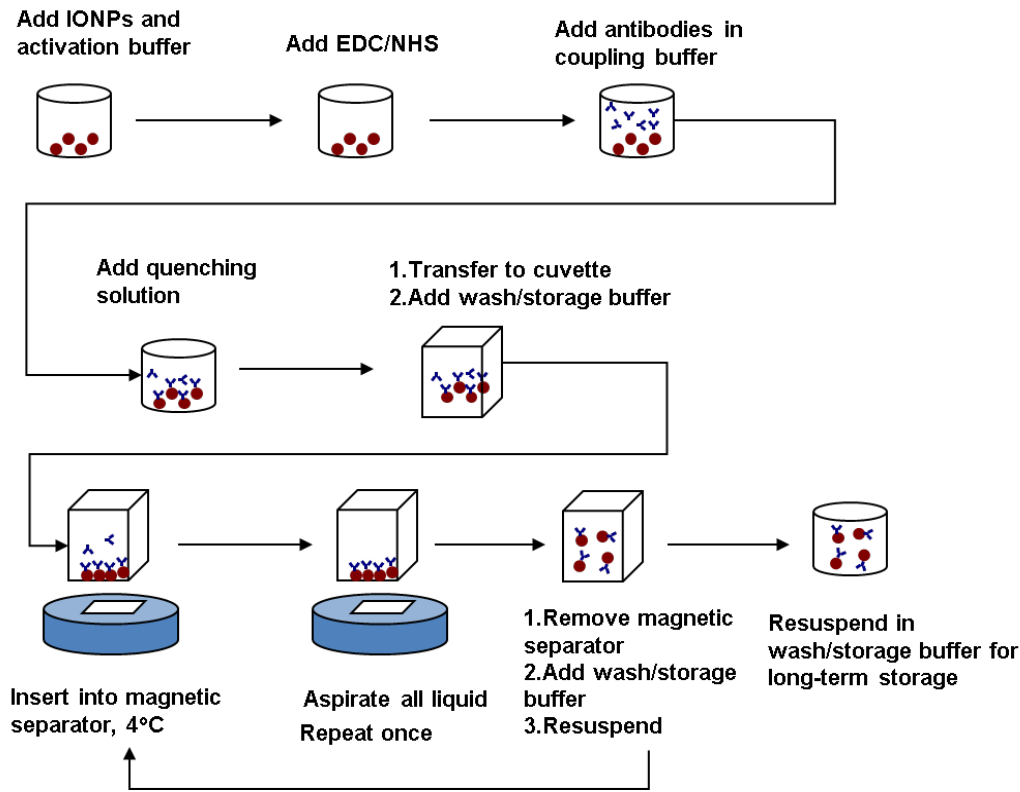

Figure 2 Schematic illustration of the preparation of detecting antibody conjugated magnetic iron oxide nanoparticles (IONPs).

### 1.1.2. Preparation of surface-biofunctionalized MR sensor

MR sensor is utilized in this platform to detect the magnetic signal of magnetic biolabels, including single MTJ sensor and MTJs array sensor (both purchased from Micro Magnetics), of which the surfaces were immobilized with capturing antibodies through a similar procedure. MTJs array sensor was bio-functionalized with anti-p24 antibodies. A thickness of 200-nm gold layer was deposited on the top of MTJ sensing area, serving as bioactive area. The characterizations of single MTJ sensor and MTJs array sensor were carried out respectively through four-point probe measurement and two-point probe measurement, as discussed in Chapter 2.4.1. The detecting antibodies were immobilized on the MTJ sensing area through the procedures as shown in Fig. 3. The key point here is adding blocking solution (5% Skim Milk in PBS with 0.05% TWEEN 20) after the incubation of MTJ sensor with antibody solution droplet. We perform this blocking step to reduce undesirable non-specific binding of protein molecules on the MTJ sensor sensing area surface which might induce false results. [10] After the immobilization of antibodies, MTJ sensor surface was washed with PBST (phosphate-buffered saline with 0.05% TWEEN-20), and the capturing antibodies bio-functionalized MTJ sensor was prepared.

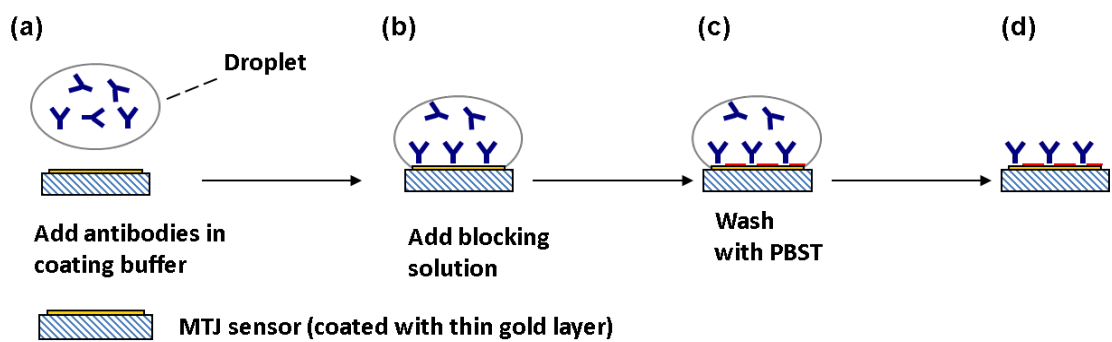

Figure 3 Schematic illustration of the preparation of capturing antibodies biofunctionalized MTJ sensor. (a) One droplet of antibody solution was introduced onto the sensor sensing area surface. (b) Capturing antibodies were passively attached to the surface of MTJ sensor by incubation in buffer. (c) Blocking buffer was added to eliminate non-specific binding effect and block the remaining active area (blocking areas are represented by red color). (d) Biofunctionalized MTJ sensor was achieved.

### 1.1.3. Magnetic immunoassay to detect biomarkers

After the biofunctionalization of the IONPs with detecting antibodies and the MTJ sensor with capturing antibodies on its sensing area, the detection of target antigen biomolecules was carried out through a sandwich immunoassay configuration as illustrated in Fig. 4. Briefly speaking, the capturing antibodies functionalized MTJ sensor sensing area was initially covered with a solution containing target antigen biomolecules. Then the sensor surface was washed with PBST after the full binding of antigens with capturing antibodies immobilized onto the MTJ sensor surface. Subsequently, the detecting antibodies functionalized IONPs were introduced, and bound to the target antigens at the different epitopes from the capturing antibodies. After removing the excess IONPs by flushing the sensor surface with PBST, a hard-axis field of 60 Oe was applied for polarizing and magnetizing the IONP. The stray magnetic field of IONPs was detected by MTJ sensor, and the signal acquisition was performed on the single MTJ sensor through four-point probe measurement, and performed on the MTJs array sensor through two-point probe measurement.

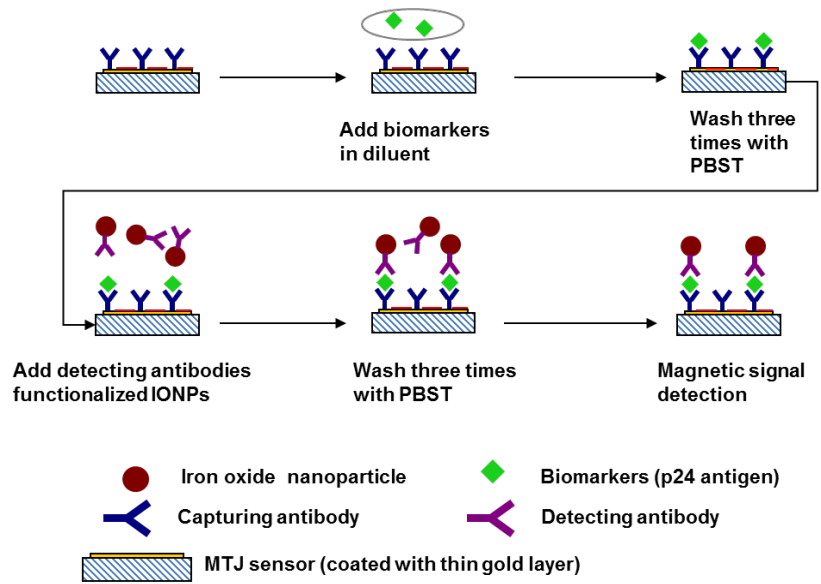

Figure 4 Schematic illustration of magnetic sandwich immunoassay based on MTJ sensor and magnetic iron oxide nanoparticles (IONPs). This system exploits the capturing antibodies functionalized MTJ sensor to capture biomarkers (antigens). The IONPs was labeled with detecting antibody. The target antigens were conjugated with both the capturing and detecting antibodies at different epitopes.

## 1.2. Results

The IONPs with carboxyl groups displayed a spherical morphology with size of 20 nm and exhibited superparamagnetic behavior with a saturation magnetization ($M_s$) of 50 emu/g.

**Detection of p24 antigens**

The MR array sensor consists of 76 junctions, and every junction stack was prepared through similar procedure as the single MTJ stack described previously. The MTJs array sensor was schematically drawn in Fig. 5a. Since the array sensor only contains two wire-bond pads, its MR transfer curve characterization was carried out through the traditional two-point probe measurement. The MTJs array sensor was placed on sample stage, shielded in a mu-metal shielding box to avoid the disturbances of ambient magnetic field. Two pairs of Helmholtz coils were utilized here. One pair of the coils provided a changeable magnetic field from -100 Oe to 100 Oe along the easy-axis and the other pair provided a 10-Oe hard-axis magnetic field on the MTJs array sensor, respectively. Fig. 5b shows the MR transfer curves of MTJs array sensor, with a sensitivity of 1.39%/Oe.

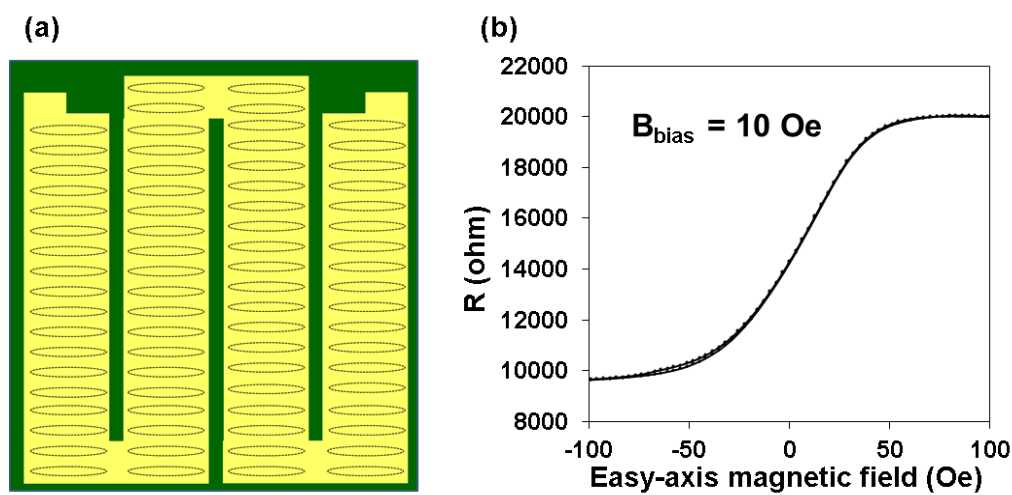

Figure 5 (a) Schematic drawing of MTJs array sensor, consisting of 76 tunneling junctions indicated by ellipses. (b) MR loop of MTJs array sensor with 10 Oe of hard-axis bias field along the hard-axis. [11]

The detection of target p24 biomolecules with concentrations of 0.001 mg/ml was realized through two-point probe measuring configuration with MTJs array sensor and biofunctionalized IONPs. A 10-Oe magnetic field was applied along the hard-axis of sensor to reduce the sensor hysteresis. The MR transfer curves of MTJs array sensor were measured before and after the binding of biofunctionalized IONPs though the conjugation with p24 on the MTJs array sensing area surface. The resistance variations ΔR of the sensor against easy-axis field changed is shown in Fig. 6a. A reference measurement was carried out where no IONPs were introduced. The maximum resistance deviations appear at approximately 50 Oe, with the maximum value of 1190 Ω, demonstrating the detection of p24 with concentration of 0.001 mg/ml. Unfortunately, it was inconvenient to reuse the same MTJs array sensor after cleaned by piranha solution as what we did to the single MTJ sensor. The main reason is due to the difference of resistance measurement methods, where single MTJ sensor was measured through four-point probe way but MTJs array sensor was measured through two-point probe way. As discussed in Chapter 2, the resistance measured with two-point probe method includes the resistance of MR sensor as well as the lead resistance and contact resistance, where the four-point probe method can make more accurate resistance measurement of MR sensor. Thus, here we demonstrated the feasibility of sensitive detection of 0.001 mg/ml p24 antigens using carboxyl-group functionalized IONPs with MTJs array sensor, where more efforts are still needed to obtain the relation curve of p24 concentration versus ΔR of MTJs array sensor.

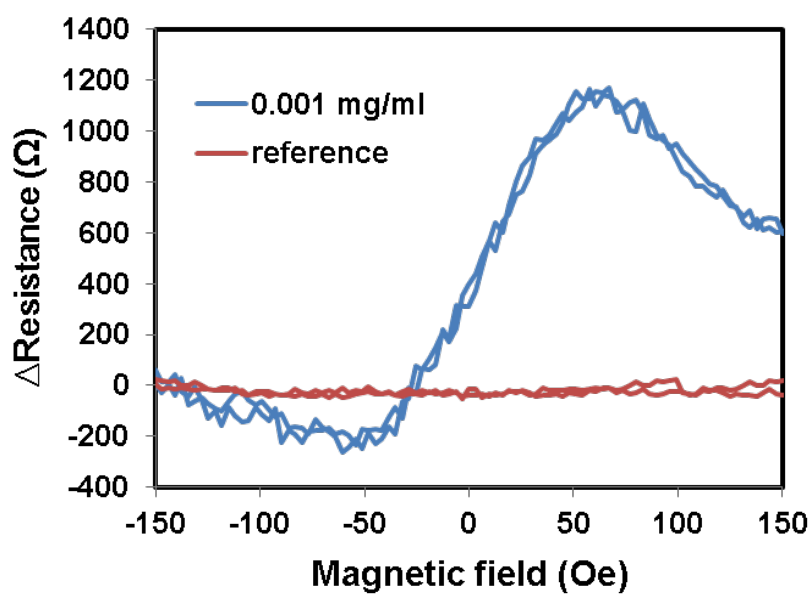

Figure 6 Diagram of the MTJ sensor resistance variation ΔR after binding with the target p24 antigens with a concentration of 0.001 mg/ml. The maximum resistance deviations appear at approximately 50 Oe, with the maximum value of 1190 Ω.

**Conclusion and outlook**

In summary, we have demonstrated the feasibility of HIV disease immunoassay with carboxyl-group functionalized IONPs and MTJ sensors by a sandwich-assay configuration. The MTJ sensors and IONPs were biologically treated with capturing antibodies and detecting antibodies before the magnetic biodetection experiments. MTJs array sensor with a sensing area of 890 ×890 μm$^2$ (almost the whole surface area of sensor) possessing a sensitivity of 1.39%/Oe was used to detect p24 antigens, HIV disease biomarker. It was demonstrated that the p24 antigens could be detected at a concentration as low as 0.001 mg/ml. However, due to the interference of contact resistance on the measured resistance through two-point probe measurement, it is inconvenient for the MTJs array sensor to be reused in new experiment after piranha wash because of varying contact resistance might be caused in every new experiment. Thus, the sensitive detection of 0.001 mg/ml p24 antigens was demonstrated here using carboxyl-group functionalized IONPs with MTJs array sensor, where more efforts are still needed to obtain the relation curve of p24 concentration versus ΔR of MTJs array sensor.

Furthermore, the magnetic biodetection of p24 antigens for health care was firstly realized here based on the configuration of MTJ sensors and carboxyl-group functionalized IONPs, where the detection sensitivity still have plenty room for improvement. It has been demonstrated by a research group in Germany that biodetection based on GMR sensor was superior to fluorescent biodetection.[12] Besides, it is known that MTJ sensor offer higher MR ratio and therefore higher sensitivity at low magnetic field than GMR sensor, and a single MTJ sensor has the ability to detect single magnetic nanoparticle.[2] Thus, we can expect that the detection sensitivity of our configuration based on MTJ sensor has potential to be improved to be less than 1 ng/ml for p24 antigens after the further optimization.

In addition, the multiplex sensing and miniaturization are also a very worthwhile future work for our magnetic biodetection configuration. The multiplex sensing of different bioanalytes can be realized by pre-mobilizing different capturing molecules on different working areas on a MR sensor array. [13] And the compatibility of MR sensor with the standard CMOS fabrication technology enables great promise for our magnetic biodetection platform to be miniaturized to for lab-on-chip application for

point-of-care with high-sensitivity and low-cost. [14]